\documentclass[12pt]{article}

\usepackage{scicite,graphicx}
\usepackage{rotating}
\usepackage{times}

\usepackage{amssymb,amsmath,bm}
\usepackage[colorlinks=true,citecolor=blue]{hyperref}

\providecommand{\apjl}{ApJ}
\providecommand{\apjs}{ApJS}

\providecommand{\aap}{A\&A}

\newenvironment{sciabstract}{%
\begin{quote} \bf}
{\end{quote}}

\title{\textbf{A clump population framework for star formation across galactic scales}}

\author{
J. W. Zhou$^{1}$,
Sami Dib$^{2}$,
Tie Liu$^{3}$,
Ralf S. Klessen$^{4,5}$,
Davide Elia$^{6}$,\\
Kee-Tae Kim$^{8,9}$,
Mika Juvela$^{7}$,
Pavel Kroupa$^{10,11}$,
Patricio Sanhueza$^{12}$,\\
Médéric Boquien$^{13}$, 
Keping Qiu$^{14}$,
Michele Ginolfi$^{15,16}$, \\
Dario Colombo$^{17}$,
Timothy Davis$^{18}$
}

\begin{document}


\baselineskip24pt


\maketitle 
\noindent \normalsize{$^{1}$Max-Planck-Institut f\"{u}r Radioastronomie, Auf dem H\"{u}gel 69, 53121 Bonn, Germany}\\
\normalsize{$^{2}$Max-Planck-Institut f\"{u}r Astronomie, K\"{o}nigstuhl 17, 69117 Heidelberg, Germany}\\
\normalsize{$^{3}$State Key Laboratory of Radio Astronomy and Technology, Shanghai Astronomical Observatory, Chinese Academy of Sciences, 80 Nandan Road, Shanghai 200030, People's Republic of China}\\
\normalsize{$^{4}$Universitat Heidelberg, Zentrum f\"{u}r Astronomie, Institut f\"{u}r
Theoretische Astrophysik, Albert-Ueberle-Str. 2, 69120, Heidelberg, Germany}\\
\normalsize{$^{5}$Universitat Heidelberg, Interdisziplin\"{a}res Zentrum f\"{u}r
Wissenschaftliches Rechnen, Im Neuenheimer Feld 205, 69120, Heidelberg, Germany}\\
\normalsize{$^{6}$INAF-Istituto di Astrofisica e Planetologia Spaziali, Via Fosso del Cavaliere 100, 00133, Roma, Italy}\\
\normalsize{$^{7}$Department of Physics, PO Box 64, 00014 University of Helsinki, Helsinki, Finland}\\
\normalsize{$^{8}$Korea Astronomy and Space Science Institute, 776 Daedeokdae-ro, Yuseong-gu, Daejeon 34055, Republic of Korea}\\
\normalsize{$^{9}$University of Science and Technology (UST), 217 Gajeong-ro, Yuseong-gu, Daejeon 34113, Republic of Korea}\\
\normalsize{$^{10}$Helmholtz-Institut f{\"u}r Strahlen- und Kernphysik (HISKP), Universität Bonn, Nussallee 14–16, 53115 Bonn, Germany}\\
\normalsize{$^{11}$Charles University in Prague, Faculty of Mathematics and Physics, Astronomical Institute, V Hole{\v s}ovi{\v c}k{\'a}ch 2, CZ-180 00 Praha 8, Czech Republic}\\
\normalsize{$^{12}$Department of Astronomy, School of Science, The University of Tokyo, 7-3-1 Hongo, Bunkyo, Tokyo 113-0033, Japan}\\
\normalsize{$^{13}$Université Côte d’Azur, Observatoire de la Côte d’Azur, CNRS,
Laboratoire Lagrange, 06000 Nice, France}\\
\normalsize{$^{14}$School of Astronomy and Space Science, Nanjing University, 163 Xianlin Avenue, Nanjing 210023, Jiangsu, P. R. China}\\
\normalsize{$^{15}$Dipartimento di Fisica e Astronomia, Università di Firenze, Via G. Sansone 1, 50019 Sesto Fiorentino (Firenze), Italy}\\
\normalsize{$^{16}$INAF – Osservatorio Astrofisico di Arcetri, Largo E. Fermi 5, I-50125 Firenze, Italy}\\
\normalsize{$^{17}$Argelander-Institut für Astronomie, University of Bonn, Auf dem Hügel 71, 53121 Bonn, Germany}\\
\normalsize{$^{18}$Cardiff Hub for Astrophysics Research \& Technology, School of Physics \& Astronomy, Cardiff University, Queens Buildings, Cardiff CF24 3AA, UK}\\


\begin{sciabstract} 
Whether star formation across different scales and galactic environments follows universal laws remains poorly understood.
We investigate cloud-scale star formation in 30 dwarf and disk galaxies. Molecular clouds are identified via the dendrogram algorithm, with their masses ($M_{\rm H_{2}}$) and star‑formation rates (SFRs) derived from integrated CO and mid-infrared emission, respectively. All galaxies exhibit a consistent $M_{\rm H_{2}}$–SFR relation that is insensitive to galactic environment. Extending our sample to include Milky‑Way cloud substructures (down to sub-parsec) and 311 diverse nearby galaxies, we confirm that this relation persists across ten orders of magnitude in both molecular gas mass and SFR. A theoretical model where star formation proceeds within dense clumps inside molecular clouds reproduces this observational trend. The weak environmental sensitivity of clump-scale star formation naturally explains the uniform $M_{\rm H_{2}}$–SFR relation across galaxies. This framework thus provides a baseline model linking star formation in the Milky Way and external galaxies.
\end{sciabstract}

\clearpage

\section*{Dense Clumps in molecular clouds}

Galaxies act as stellar factories in the universe, forming stars through the gravitational collapse of the densest regions of molecular clouds (clumps) within the interstellar medium (ISM). Molecular clouds are widely distributed throughout galaxies and serve as the localized sites of star formation. Usually, each cloud contains multiple clumps, which represent the birthplaces of stars and the precursors of embedded star clusters \cite{McKee2007-45,Zinnecker2007-45,Kennicutt2012-50,Krumholz2014-539,Heyer2015-53,Motte2018-56,Vazquez2019-490}. 
The hierarchical structure of molecular gas plays a central role in shaping the spatial distribution of emerging stellar populations \cite{Elmegreen2014-787,Chevance2023-534}.
The typical surface density and radius of molecular clouds in the inner Milky Way are $\sim$110 M$_\odot$/pc$^{2}$ and $\sim$20 pc \cite{Roman-Duval2010-723, Heyer2015-53, bridge}, respectively, though these values depend on how molecular clouds are defined. For the clumps in the catalog of \cite{Urquhart2022-510}, the values of these two quantities are $\sim$910 M$_\odot$/pc$^{2}$ and $\sim$0.3 pc, respectively.

High-resolution observations across multiple wavelengths have now made it possible to dissect star formation down to individual molecular clouds in nearby galaxies. ALMA CO imaging enables systematic studies of cloud populations beyond the Milky Way \cite{Schinnerer2024-62}, while JWST and HST data reveal embedded clusters, young clusters, and stellar associations within those clouds \cite{Rodriguez2025-983,Thilker2025-280}. The combination of these datasets provides an unprecedented view of the connection between individual clouds and their stellar populations \cite{Turner2022-516,Zhou2025-541}. 
Molecular clouds in the Milky Way, which can easily be resolved down to internal clump scales, provide a crucial benchmark for interpreting cloud-scale star formation beyond the Milky Way. 
Earlier studies have verified that
the star formation rates of molecular clouds are mainly driven by their dense gas content \cite{Gao2004-606,Wu2005-635,Lada2010-724,Lada2012-745}. 
In this work, we further specify that dense gas refers to star-forming clumps and regard these clumps as the fundamental units of star formation, upon which our theoretical framework is built. 
Clumps are gravitationally bound, dense substructures within molecular clouds and represent the observationally established sites where stars form \cite{Motte2018-56,Urquhart2022-510,Zhou2024PASP-2}. By describing clumps as the fundamental units of star formation, star formation activity observed on larger scales, such as molecular clouds and kpc-scale molecular gas complexes, arises from the collective contribution of their internal clump populations. For example, the star formation rate of a molecular cloud can be estimated by summing the contributions from its constituent star-forming clumps.
Understanding star formation on large scales therefore requires characterizing the physical and evolutionary states of clumps. Surveys such as ATLASGAL and Hi-GAL \cite{Schuller2009-504,Urquhart2022-510,Molinari2010,Elia2021-504}, together with targeted ALMA follow-up observations \cite{Sanhueza2019-886,Liu2020,Motte2022-662,Liu2024-24,Molinari2025-696}, provide the necessary foundation.

Comparisons of star formation across different scales and types of systems have been discussed in previous studies \cite{Lada2012-745,Zhang2014-784,Liu2016-829,Nguyen2016-833,Stephens2016-824,Jimenez2019-880,Neumann2025-693}. Thanks to recent observational advances, however, our sample is substantially more extensive than those used in earlier works, and the clump catalogs that form the basis of our model have only become available in recent years \cite{Elia2021-504,Urquhart2022-510}. These new datasets enable a unified analysis spanning nearly ten orders of magnitude in both molecular gas mass and star formation rate.
This work studies star formation on molecular-cloud scales across 30 dwarf and disk galaxies, spanning a wide range of galactic environments and molecular cloud types (see Table.S\ref{near} and Fig.S\ref{s-para}), to investigate the cross-galaxy consistency of cloud-scale star formation. The effectiveness of a unified theoretical framework for star formation on molecular-cloud scales across different galaxies is also explored.

\section*{Cross-galaxy $M_{\rm H_{2}}$--SFR relation}

\begin{figure*}
\centering
\includegraphics[width=1\textwidth]{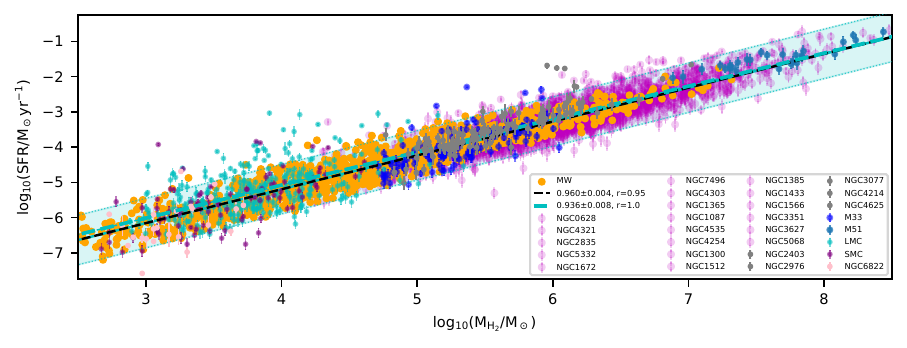}

\vspace{-15pt}
\caption{The $M_{\rm H_{2}}$--SFR relation for molecular clouds in 30 galaxies, see also Fig.S\ref{s-all} for more details. Error bars represent the uncertainty in the SFR estimates. 
The black dashed line shows the linear fit (unweighted ordinary least squares) to all data points, while the cyan dashed line represents the model prediction (see Sec.\ref{procedure} for more details). $r$ is the correlation coefficient.
The 95\% prediction bands are shown as shaded regions with dashed boundary lines.
}
\label{msfr}
\end{figure*}

The calculations of molecular cloud mass and star formation rate are described in detail in Sec.\ref{calculation}.
Fig.\ref{msfr} shows that molecular clouds in all 30 galaxies follow a similar $M_{\rm H_{2}}$--SFR relation, except for a subset of molecular clouds in dwarf galaxies that show elevated SFRs, see Sec.\ref{h-sfr} for more discussion. 

Molecular cloud properties vary across regions within a galaxy and also from galaxy to galaxy \cite{Querejeta2021-656,Kim2022-516,Sun2022-164}. However,
as discussed in Sec.\ref{environment}, the $M_{\rm H_{2}}$--SFR relation shows no significant dependence on the galactic environment.
This further confirms that molecular clouds are not the fundamental units of star formation, but that the true sites of star formation are the locally dense clumps within them.
The stability of the $M_{\rm H_{2}}$--SFR relation across different galaxies and environments reflects the relative independence of clump-scale star formation. 
This independence is a natural consequence of gravitational collapse: the highest-density peaks have the shortest dynamical timescales, causing them to collapse and form stars first. 
These clumps can therefore be regarded as mutually independent nodes within molecular clouds \cite{Vazquez2019-490,Zhou2025-699}. This independence is further supported by the suggestion that stellar feedback does not significantly alter the physical properties of the surrounding dense gas structures \cite{Watkins2019-628A, Zhou2024-682-128, Zhou2024-682-173}, though these studies are limited to specific regions of the Milky Way. For more extreme environments, dense clumps can survive in the vicinity of R136 in 30 Doradus \cite{Valdivia2025-699}.
At the spatial scale of $\sim$0.2 pc, the star formation efficiency (SFE) of dense clumps in the Central Molecular Zone is comparable to that in Galactic disk sources \cite{Lu2019-872}. In addition, while the SFE of clumps can be a few tens of percent ($\sim$0.3) \cite{Zhou2024PASP-2}, the SFE of molecular clouds is only a few percent \cite{Chevance2023-534}. The low SFE of molecular clouds essentially reflects an average over high-SFE, star-forming clumps and a substantial reservoir of diffuse, non–star-forming gas.

In Sec.\ref{addition}, by incorporating molecular cloud substructures in the Milky Way down to sub-parsec scales, together with 311 nearby galaxies of diverse types, we construct a sample spanning ten orders of magnitude in both mass and SFR, and find that the $M_{\rm H_{2}}$--SFR relation still holds. 
This is consistent with a hierarchical view in which cloud-scale star formation emerges from the collective activity of clump populations, since all the above structures are larger-scale systems containing internal clumps.
Moreover, this multi-scale consistency suggests the possibility of a unified theory of star formation in the Milky Way and nearby galaxies. 
While large-scale structures exhibit significant diversity, their star formation properties show remarkable consistency when described through the underlying populations of star-forming clumps.

\section*{Theoretical model}\label{model}

The molecular clouds we observe are typically composites, containing both clumps and stellar populations, whether exposed or still embedded. Almost all of these stellar populations originate from clumps, which may currently exist or have existed in the past. 
Together, the existing and past clumps constitute the cloud’s initial clump population, described by the initial clump mass function \cite{model}. Over time, this initial population has evolved into a mixture of presently observed young clumps and the embedded/exposed stellar populations. Young clumps are in earlier evolutionary stages, whereas embedded and exposed stellar populations trace progressively later stages following star formation and gas dispersal. Embedded stellar populations are associated with more evolved clumps in current observations.
In this sense, the different components of a molecular cloud — young clumps, embedded, and exposed stellar populations — trace the age spread of clumps within the cloud. Based on this picture, we constructed a model \cite{model} that integrates observational results from the Milky Way and nearby galaxies to predict cloud-scale SFRs.
In the model, we first estimate the cloud’s total initial clump mass based on its overall mass, and then generate the initial clump population using the initial clump mass function. Next, we model the cloud’s star formation history (SFH) to assign ages to each clump. The intermediate-age clumps are then sorted, and the total mass of embedded clusters is calculated. Finally, the SFR is predicted based on the duration of the embedded phase.

Beyond the dense gas fraction extensively discussed in previous works, our framework explicitly accounts for the evolutionary states of individual star-forming clumps. At any given observational snapshot, only a subset of dense clumps within a molecular cloud are actively forming stars, whereas others remain quiescent despite having similarly high densities. This information is inaccessible when dense gas is treated as a single bulk component, but becomes available once it is resolved into a population of individual clumps. Consequently, the dense gas fraction alone is insufficient to explain the observed differences in star formation rates among molecular clouds of similar mass, whereas the evolutionary distribution of clumps naturally contributes to the observed scatter, although it is not necessarily the only factor responsible for the scatter.

The procedure for fitting the observations with this model can be referred to \cite{model}, the online code, and the overview provided in Sec.\ref{procedure}.
In Fig.\ref{msfr}, the model can fit the observations in terms of slope and intercept, although it is slightly overestimated (see Sec.\ref{procedure} for more details).
The agreement between the baseline model and observations demonstrates its effectiveness across molecular clouds in different galaxies, providing a framework for a unified theory of star formation across molecular cloud and larger galactic scales.
Moreover, the consistency of the $M_{\rm H_{2}}$--SFR relation across galaxies indicates that molecular clouds in the Milky Way can serve as useful templates for studying molecular clouds in nearby galaxies.
In the Milky Way, we are able to investigate in detail the underlying physical processes within individual molecular clouds, as well as how these small-scale processes combine to produce the observed star formation phenomena on cloud and larger scales. This is important for studies of star formation in nearby galaxies, where even with the most advanced observational facilities (such as ALMA and JWST), we are still limited to resolving structures at the scale of molecular clouds. However, a description based solely on large-scale, macroscopic properties does not fully capture the essential physics of star formation. Treating dense gas as a population of individual clumps allows the model to incorporate the mass distribution and evolutionary state of clump populations, rather than describing the cloud solely through a single dense gas fraction. This provides a natural framework for integrating the rapidly growing body of high-resolution Galactic ALMA observations, which increasingly focus on the physical properties and evolution of individual clumps. We therefore regard the present work as a step toward connecting these detailed Galactic studies with the large-scale star formation relations observed in nearby galaxies.

Finally, we note that our framework does not address the fundamental physical origin of clumps within molecular clouds. Importantly, the near-linear $M_{\mathrm{H_2}}$--SFR relation itself is not a discriminating prediction among most star formation frameworks, as the same near-linear relation can arise in different physical scenarios, including turbulence-regulated, gravity-driven, feedback-regulated, or combined processes \cite{Krumholz2005-630,Federrath2012-761,Dib2011-737,Grudic2019-488,Ballesteros2024-534}. Testing the underlying theoretical models for dense clump formation requires more detailed constraints on gas flows and energetics of their small-scale substructures from systematic high-resolution observations \cite{2026MNRAS.547f2059V}.

\bibliography{ref}
\bibliographystyle{Science}
{\bf Data and materials availability:} 
The implementation code and its documentation are available online
\footnote{\url{https://github.com/jianwenzhou11/SFR_model.git}}.

For nearby galaxies, all the data used in this work are available on the PHANGS team website
\footnote{\url{https://sites.google.com/view/phangs/home}}. For the Milky Way, the datasets are available on the team websites of the ThrUMMS \footnote{\url{https://gemelli.spacescience.org/~pbarnes/research/thrumms/rbank/}}, the FUGIN \footnote{\url{https://nro-fugin.github.io/release/}}, the MWISP \footnote{\url{https://doi.org/10.57760/sciencedb.27351}}  and the HERACLES \footnote{\url{https://iram-institute.org/science-portal/proposals/lp/completed/lp001-the-hera-co-line-extragalactic-survey/}} surveys. 
The source identification package {\tt astrodendro} is available at \url{http://dendrograms.org/}. 

Correspondence and requests for materials should be addressed to Jianwen Zhou
(jianwenzhou11@gmail.com).

{\bf Funding:}
T.L. acknowledges support from the National Key R\&D Program of China (No. 2022YFA1603100), the National Science and Technology Major Project of China (No. 2024ZD1100601), the National Natural Science Foundation of China (NSFC), through grants No. 12073061 and No. 12122307, the Tianchi Talent Program of Xin-jiang Uygur Autonomous Region, and the international partnership program of the Chinese Academy of Sciences, through grant No. 114231KYSB20200009. 
R.S.K. acknowledges financial support from the ERC via Synergy Grant ``ECOGAL'' (project ID 855130) and from the German Excellence Strategy via the Heidelberg Cluster ``STRUCTURES'' (EXC 2181 - 390900948). In addition RSK is grateful for funding from the German Ministry for Economy and Energy (BMWE) in project ``MAINN'' (funding ID 50OO2206), and from DFG and ANR for project ``STARCLUSTERS'' (funding ID KL 1358/22-1). 
M.J. acknowledges the support of the Research Council of Finland Grant No. 348342.
P.K. acknowledges support through the DAAD Eastern European Exchange Programme between Bonn and Charles Universities and through Hotel Castle Residence.

\section*{Acknowledgments}
We thank the referees for their helpful and constructive comments that improved the quality of this work.
We thank Amélie Saintonge, Neal Evans, Leslie Hunt, Guangxing Li and Di Li for the helpful comments.
It is a pleasure to thank the PHANGS team; the data cubes and other data products they shared made this work possible and easy to carry out.
ALMA is a partnership of ESO (representing its member states), NSF (USA) and NINS (Japan), together with NRC (Canada), NSTC and ASIAA (Taiwan), and KASI (Republic of Korea), in cooperation with the Republic of Chile. The Joint ALMA Observatory is operated by ESO, AUI/NRAO and NAOJ. 
This work is based on observations made with the NASA/ESA/CSA James Webb Space Telescope.
This research made use of data products from the Midcourse Space Experiment (MSX). Processing of the data was funded by the Ballistic Missile Defense Organization with additional support from NASA Office of Space Science. This research has also made use of the NASA/ IPAC Infrared Science Archive, which is operated by the Jet Propulsion Laboratory, California Institute of Technology, under contract with the National Aeronautics and Space Administration.

{\bf Authors contributions:}
J.W.Z. led the data analysis and interpretation, and wrote major sections of the main text and Methods. S.D. and P.K. contributed to the development and application of the theoretical models and their interpretation of the observational results. They also contributed to the revision of the manuscript. R.S.K., T.L., D.C., D.E., M.B., P.S., T.D., M.G., K.-T.K., K.Q. and M.J. reviewed the manuscript and provided comments that improved its quality. R.S.K. and S.D. provided further clarification of the theoretical framework presented in this work.

{\bf Competing interests:} There are no competing interests to declare. 

\clearpage

%

\renewcommand{\figurename}{Figure~S}
\renewcommand{\tablename}{Table~S}
\makeatletter
\def\fnum@figure{\figurename\thefigure}
\def\fnum@table{\tablename\thetable}
\makeatother
\setcounter{figure}{0}
\setcounter{table}{0}

\section*{Methods}
\begin{sidewaystable}
\centering
\caption{Physical parameters of 30 galaxies—from left to right: 
linear resolution of CO observations, linear resolution of mid-infrared observations, distance, position angle, inclination, molecular cloud number, CO observations, molecular hydrogen mass, star formation rate, stellar mass, metallicity, and references for galaxy parameters. For CO observations, "1" \cite{Leroy2021-257,Leroy2021-255,Lee2023-944,Williams2024-273}, "2" \cite{Wong2011-197,Wong2017-850}, "3" \cite{Gratier2010-522,Druard2014-567}, "4" \cite{denBrok2022-662}, "5" \cite{Barnes2015-812, Barnes2025-280, Umemoto2017-69},
"6" \cite{Leroy2009-137,Schruba2012-143}, "7" \cite{Schruba2017-835}, "8" \cite{Saldano2023-672}.} 
\vspace{0.5em}
\small
\begin{tabular}{ccccccccccccc}
\hline
galaxies & linear-CO & linear-MIR & $D$ & P. A. & $i$ & clouds & obs. & $\log M_{\rm H_{2}}$ & $\log {\rm SFR}$ & $\log M_{*}$ & $12+\log {\rm O/H}$ & ref. \\
& (pc) & (pc) & (Mpc) & (deg) & (deg) & & & $(M_\odot)$ & $(M_\odot/{\rm yr})$ & $(M_\odot)$ & & \\     
\hline   

IC 5332	&	32	&	31	&	9	&	74	&	27	&	9	&	1	&	7.92 	&	-0.39 	&	9.68 	&	8.52 	&	\cite{Kreckel2019-887,Leroy2021-257}	\\
NGC 0628	&	54	&	33	&	9.8	&	21	&	9	&	732	&	1	&	9.28 	&	0.24 	&	10.34 	&	8.69 	&	\cite{Schruba2012-143,Leroy2021-257}	\\
NGC 1087	&	123	&	54	&	15.9	&	359	&	43	&	167	&	1	&	9.15 	&	0.11 	&	9.94 	&	8.53 	&	\cite{Kreckel2019-887,Leroy2021-257}	\\
NGC 1300	&	113	&	64	&	19	&	278	&	32	&	254	&	1	&	9.33 	&	0.07 	&	10.62 	&	8.52 	&	\cite{Barnes2026-706,Leroy2021-257}	\\
NGC 1365	&	131	&	66	&	19.6	&	201	&	55	&	198	&	1	&	10.32 	&	1.24 	&	11.00 	&	8.64 	&	\cite{Pilyugin2014-147,Leroy2021-257,delosReyes2019-872}	\\
NGC 1385	&	106	&	58	&	17.2	&	181	&	44	&	186	&	1	&	9.20 	&	0.32 	&	9.98 	&	8.45 	&	\cite{Leroy2021-257,delosReyes2019-872}	\\
NGC 1433	&	99	&	63	&	18.6	&	200	&	29	&	187	&	1	&	9.30 	&	0.05 	&	10.87 	&	8.54 	&	\cite{Barnes2026-706,Leroy2021-257}	\\
NGC 1512	&	95	&	64	&	18.8	&	262	&	43	&	160	&	1	&	9.09 	&	0.11 	&	10.72 	&	8.56 	&	\cite{Leroy2021-257,Hunt2019-621}	\\
NGC 1566	&	108	&	60	&	17.7	&	215	&	30	&	588	&	1	&	9.72 	&	0.66 	&	10.79 	&	8.55 	&	\cite{Barnes2026-706,Leroy2021-257}	\\
NGC 1672	&	182	&	66	&	19.4	&	134	&	43	&	173	&	1	&	9.88 	&	0.88 	&	10.73 	&	8.57 	&	\cite{Kreckel2019-887,Leroy2021-257}	\\
NGC 2835	&	50	&	41	&	12.2	&	1	&	41	&	130	&	1	&	8.54 	&	0.10 	&	10.00 	&	8.57 	&	\cite{Kreckel2019-887,Leroy2021-257}	\\
NGC 3351	&	71	&	34	&	10	&	193	&	45	&	271	&	1	&	8.96 	&	0.12 	&	10.37 	&	8.90 	&	\cite{Schruba2012-143,Leroy2021-257}	\\
NGC 3627	&	89	&	38	&	11.3	&	173	&	57	&	471	&	1	&	9.81 	&	0.59 	&	10.84 	&	8.66 	&	\cite{Schruba2012-143,Leroy2021-257}	\\
NGC 4254	&	113	&	44	&	13.1	&	68	&	34	&	480	&	1	&	9.76 	&	0.49 	&	10.42 	&	8.79 	&	\cite{Schruba2012-143,Leroy2021-257}	\\
NGC 4303	&	149	&	58	&	17	&	312	&	24	&	334	&	1	&	9.83 	&	0.73 	&	10.51 	&	8.57 	&	\cite{Leroy2021-257,delosReyes2019-872}	\\
NGC 4321	&	123	&	52	&	15.2	&	156	&	39	&	583	&	1	&	9.85 	&	0.55 	&	10.75 	&	8.50 	&	\cite{delosReyes2019-872,Leroy2021-257}	\\
NGC 4535	&	119	&	53	&	15.8	&	180	&	45	&	330	&	1	&	9.44 	&	0.34 	&	10.54 	&	8.47 	&	\cite{Kreckel2019-887,Leroy2021-257}	\\
NGC 5068	&	26	&	18	&	5.2	&	342	&	36	&	318	&	1	&	8.09 	&	-0.56 	&	9.41 	&	8.57 	&	\cite{Pilyugin2014-147,Leroy2021-257}	\\
NGC 7496	&	152	&	63	&	18.7	&	194	&	36	&	155	&	1	&	9.16 	&	0.35 	&	10.00 	&	8.49 	&	\cite{Barnes2026-706,Leroy2021-257}	\\
LMC	&	11	&	1.5	&	0.05	&	180	&	30	&	686	&	2	&	7.93 	&	-0.70 	&	9.18 	&	8.30 	&	\cite{Schruba2012-143,McConnachie2012-144}	\\
M33	&	49	&	24	&	0.84	&	201	&	54	&	215	&	3	&	8.24 	&	-1.04 	&	9.55 	&	8.40 	&	\cite{delosReyes2019-872}	\\
MW	&	--	&	--	&	--	&	--	&	--	&	1304	&	5	&	9.30 	&	0.18 	&	10.80 	&	8.69 	&	\cite{Licquia2015-806,Heyer2015-53}	\\
M51	&	472	&	221	&	7.6	&	170	&	20	&	29	&	4	&	9.66 	&	0.38 	&	10.82 	&	8.55 	&	\cite{delosReyes2019-872}	\\
NGC 2403	&	201	&	93	&	3.2	&	124	&	63	&	85	&	6	&	8.44 	&	0.31 	&	9.86 	&	8.57 	&	\cite{Schruba2012-143}	\\
NGC 2976	&	226	&	104	&	3.6	&	335	&	65	&	16	&	6	&	8.00 	&	0.09 	&	9.27 	&	8.67 	&	\cite{Schruba2012-143,delosReyes2019-872}	\\
NGC 3077	&	239	&	110	&	3.8	&	45	&	46	&	3	&	6	&	7.45 	&	0.07 	&	9.17 	&	8.64 	&	\cite{Schruba2012-143,Leroy2019-244}	\\
NGC 4214	&	182	&	84	&	2.9	&	65	&	44	&	5	&	6	&	8.01 	&	0.10 	&	8.95 	&	8.25 	&	\cite{Schruba2012-143,delosReyes2019-872}	\\
NGC 4625	&	598	&	276	&	9.5	&	330	&	47	&	4	&	6	&	7.37 	&	-1.56 	&	9.19 	&	8.35 	&	\cite{delosReyes2019-872}	\\
NGC 6822	&	2	&	13	&	0.47	&	115	&	60	&	156	&	7	&	6.88 	&	-1.85 	&	8.32 	&	8.02 	&	\cite{Schruba2012-143,delosReyes2019-872}	\\
SMC	&	9	&	1.7	&	0.06	&	155	&	64	&	177	&	8	&	7.10 	&	-1.30 	&	8.66 	&	8.00 	&	\cite{Schruba2012-143,McConnachie2012-144}	\\
\hline
\label{near}
\end{tabular}
\end{sidewaystable}

\section{Data and calculation}\label{calculation}
To ensure comparable resolution between the CO and mid-infrared observations, we used ALMA CO data and JWST 21~$\mu$m data from the PHANGS  (Physics at High Angular resolution in Nearby GalaxieS) survey for 19 nearby galaxies \cite{Leroy2021-257,Lee2023-944}. For the Milky Way (MW), Large Magellanic Cloud (LMC), M33, M51, NGC 2403, NGC 2976, NGC 3077, NGC 4214, and NGC 4625, we used single-dish CO observations combined with \textit{Spitzer} 24~$\mu$m data \cite{Dale2009-703,Kennicutt2003-115}. Additional information about these galaxies are listed in Table.S\ref{near}. Except for the Milky Way and the LMC, where CO observations correspond to the $J=1\!-\!0$ transition, other galaxies are observed in the $J=2\!-\!1$ transition.For NGC 6822 and the Small Magellanic Cloud (SMC), we directly use the cloud catalogs from \cite{Schruba2017-835, Saldano2023-672}.

In \cite{bridge,Zhou2024PASA,Zhou2024-534,Zhou2025-541}, 
we identified molecular clouds and discussed their physical properties in detail in the Milky Way and nine nearby galaxies.
For the additional galaxies included in the sample, we identify molecular clouds using the same method presented in previous work by the dendrogram algorithm. 
As described in \cite{Rosolowsky2008-679}, the dendrogram algorithm decomposes density or intensity data into hierarchical structures called leaves, branches, and trunks. 
In the {\it astrodendro} package\footnote{\url{https://dendrograms.readthedocs.io/en/stable/index.html}},
there are three major input parameters for the dendrogram algorithm: {\it min\_value} for the minimum value to be considered in the dataset, {\it min\_delta} for a leaf that can be considered as an independent entity, and {\it min\_npix} for the minimum area of a structure.
Given that all retained structures in the strictly masked Moment 0 map are considered reliable, we only require that the smallest identified structure exceeds one beam area. No additional constraints are imposed within the algorithm to reduce the dependence of the identification results on parameter choices. The number of identified molecular clouds (leaves) in each galaxy is shown in Table.S\ref{near}.

We use the following formulas to convert CO integrated intensity ($I_{\rm CO}$) into molecular cloud mass, and mid-infrared emission ($I_{\rm MIR}$) into star formation rate \cite{Sun2022-164}:
\begin{equation}
    \frac{\Sigma_{\rm SFR}}{\rm M_{\odot}~yr^{-1}~kpc^{-2}} = 3.8\times10^{-3} \left(\frac{I_{\rm MIR}}{\rm MJy~sr^{-1}}\right),
\end{equation}
\begin{equation}
    \mathrm{SFR} = \Sigma_{\rm SFR} \times A,
    \label{21um}
\end{equation}
\begin{equation}
M_{\rm H_{2}} = \alpha_{\rm CO} \times I_{\rm CO} \times A,
\label{mass}
\end{equation}
where $A$ is the physical area of the molecular cloud, $\alpha^{2-1}_{\rm CO} \approx 6.7~\rm M_{\odot} \rm pc^{-2} (\rm K~km~s^{-1})^{-1}$ and $\alpha^{1-0}_{\rm CO} \approx 4.35~\rm M_{\odot} \rm pc^{-2} (\rm K~km~s^{-1})^{-1}$ are the CO-to-H$_{2}$ conversion factors \cite{Leroy2022-927}. 
Here, we adopt the simplest possible prescriptions to estimate the molecular cloud mass and star formation rate, in order to minimize the introduction of uncertainties and maintain consistency in the measurements across samples.

Molecular gas and star formation activity closely track each other when observations are averaged over individual regions—yielding a molecular Kennicutt–Schmidt relation in which the surface densities of star formation rate ($\Sigma_{\rm SFR}$) and molecular gas mass ($\Sigma_{\rm mol}$) correlate well on galactic or (sub-)kiloparsec scales \cite{Kennicutt1998-498, Bigiel2008-136, Schinnerer2024-62}. 
However, on the molecular cloud scale, both $\Sigma_{\rm SFR}$ and $\Sigma_{\rm mol}$ can be significantly affected by the cloud size, and are thus subject to beam effects, which become particularly severe when comparing molecular clouds across different galaxies. To minimize these biases, we directly study the total SFR and total molecular mass derived from integrated fluxes (i.e., the $M_{\rm H_{2}}$--SFR relation) rather than using surface densities. 
For extragalactic observations, considering that differences in resolution or the evolution of molecular clouds may cause offsets between CO and mid-infrared emission \cite{Chevance2020-493,Kim2022-516}, we estimate the SFR of each molecular cloud within the radii of one and 1.5 times the cloud size, representing the lower and upper bounds, respectively.

To ensure that the mid-infrared emission is strictly associated with molecular clouds, we calculate the ratio ($r_{\rm m}$) between the average mid-infrared emission within the molecular cloud and the one in an annular region outside the cloud with a width equal to the cloud’s radius. We also compute the ratio ($r_{\rm p}$) between the peak mid-infrared emission within the cloud and the peak emission in the same annular region. We require both $r_{\rm m}$ and $r_{\rm p}$ to be greater than 1.
This criterion excludes overly evolved molecular clouds, where stellar feedback has disrupted the original cloud, leaving only fragments such that the remaining molecular gas no longer represents the original cloud. 

\section{Additional samples}\label{addition}

\begin{figure*}
\centering
\includegraphics[page=1,width=0.9\textwidth]{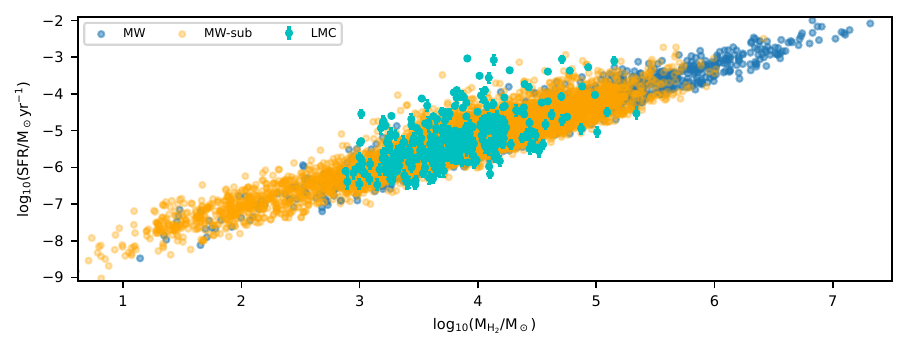}

\vspace{-25pt}
\includegraphics[page=2,width=0.9\textwidth]{fig/s-multi.pdf}

\vspace{-15pt}
\caption{
The $M_{\rm H_{2}}$--SFR relation from sub-pc to kpc scales.
(a) Blue and orange points represent molecular clouds and their substructures in the Milky Way, respectively. Cyan dots with error bars show molecular clouds in the LMC. Error bars represent the uncertainty in the SFR estimates;
(b) Blue and orange points represent gas structures (molecular clouds or molecular gas complexes) in five nearby galaxies (NGC 0628, NGC 4321, NGC 4254, NGC 3351 and NGC 3627) identified from low-resolution single-dish observations and high-resolution ALMA observations, respectively.
}
\label{s-multi}
\end{figure*}

\begin{figure*}
\centering
\includegraphics[width=0.95\textwidth]{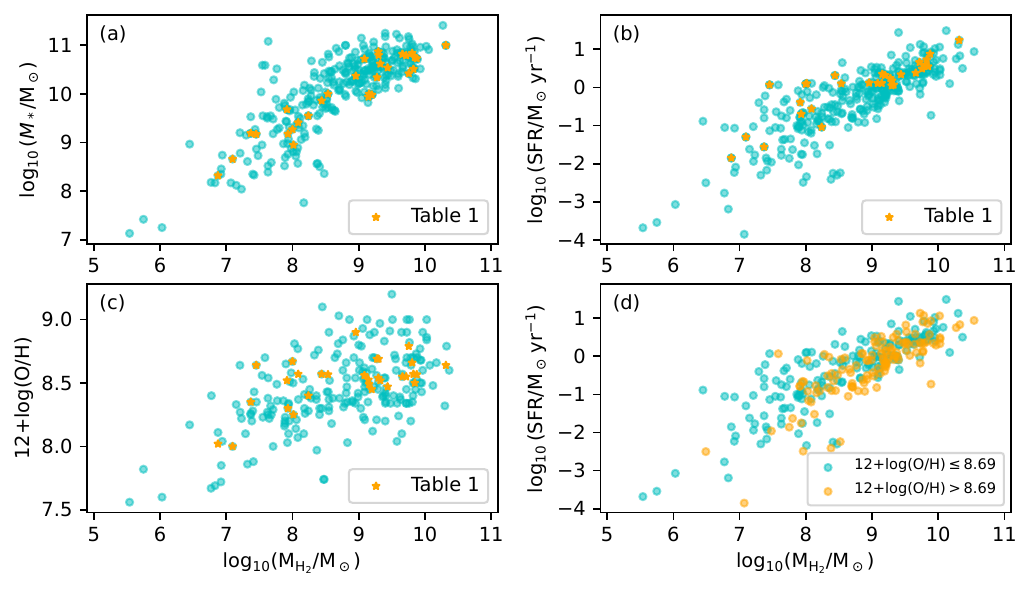}

\vspace{-15pt}
\caption{
The dependence of stellar mass, metallicity, star formation rate, and molecular gas mass of galaxies. Orange star symbols represent the galaxies listed in Table.S\ref{near}.
}
\label{s-para}
\end{figure*}


\begin{figure*}
\centering
\includegraphics[page=1,width=0.9\textwidth]{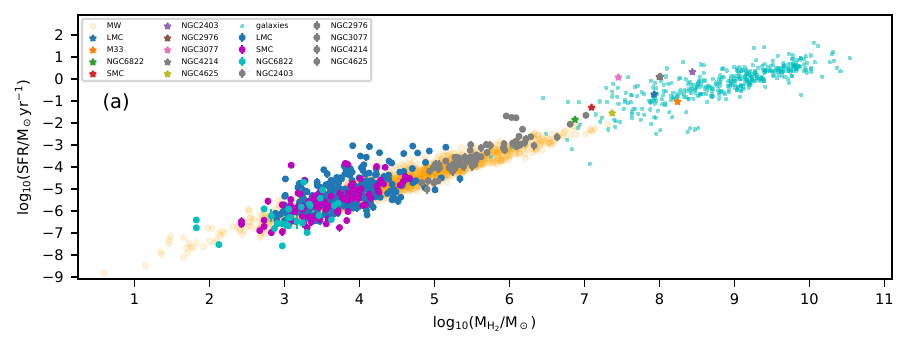}

\vspace{-25pt}
\includegraphics[page=2,width=0.9\textwidth]{fig/s-all.pdf}

\vspace{-24pt}
\includegraphics[page=3,width=0.9\textwidth]{fig/s-all.pdf}

\vspace{-15pt}
\caption{
The $M_{\rm H_{2}}$--SFR relation for all samples.
The cyan squares represent 311 nearby galaxies, considering only their global properties. The star symbols indicate the galaxies for which the internal molecular cloud populations are further investigated (listed in Table.S\ref{near}).
(a) Background orange points represent molecular clouds in the Milky Way. 
Colored dots with error bars show molecular clouds in different dwarf galaxies, which are also indicated by colored star symbols. Error bars represent the uncertainty in the SFR estimates; 
(b) Background orange and blue points represent molecular clouds and their substructures in the Milky Way, respectively. Colored dots with error bars show molecular clouds in other disk galaxies listed in Table.S\ref{near}. Colored star symbols mark all disk galaxies. 
As in Fig.S\ref{s-multi}, "single-dish" represents molecular gas structures (molecular clouds or molecular gas complexes) identified from low-resolution single-dish observations in five nearby galaxies (NGC 0628, NGC 4321, NGC 4254, NGC 3351, and NGC 3627);
(c) Combining panels (a) and (b), orange and blue dots represent disk and dwarf galaxies, respectively, which are also indicated by colored star symbols.
}
\label{s-all}
\end{figure*}

\begin{figure*}
\centering
\includegraphics[page=1,width=0.9\textwidth]{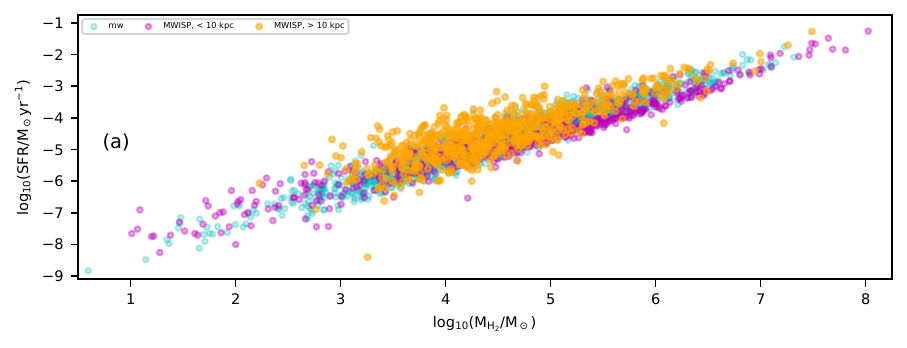}

\vspace{-25pt}
\includegraphics[page=2,width=0.9\textwidth]{fig/s-outer.pdf}

\vspace{-15pt}
\caption{
The $M_{\rm H_{2}}$--SFR relation for molecular clouds in the outer Milky Way.
(a) Cyan points denote Milky Way molecular clouds listed in Table.S\ref{near}. Magenta and orange points correspond to molecular clouds identified from the MWISP $^{12}$CO (1–0) survey data, with the former located at $R_{\rm GC} <$ 10 kpc and the latter at $R_{\rm GC} >$ 10 kpc. $R_{\rm GC}$ is the Galactocentric distance;
(b) Same as Fig.S\ref{s-all}(a), but with the Milky Way molecular clouds listed in Table.S\ref{near} replaced by molecular clouds identified from the MWISP $^{12}$CO (1–0) survey data in Sec.\ref{outer}. 
}
\label{s-outer}
\end{figure*}

Based on Table.S\ref{near},
we further incorporate a variety of additional samples spanning different galaxy types, thereby significantly increasing the diversity of the dataset and covering a wide range of the parameter space.

\subsection{Molecular gas on sub-pc and kpc scales}\label{sub-pc}

In \cite{bridge}, 
based on the $^{12}$CO (1-0) data from the ThrUMMS (Three-mm Ultimate Mopra Milky Way Survey, $300^\circ \le l \le 360^\circ$ and $-1^\circ \le b \le 1^\circ$)  and FUGIN (FOREST Unbiased Galactic plane Imaging survey with the Nobeyama 45 m telescope, $10^\circ \le l \le 50^\circ$ and $-1^\circ \le b \le 1^\circ$) surveys \cite{Barnes2015-812, Barnes2025-280, Umemoto2017-69}, we identified 1304 molecular clouds in the inner Milky Way.
We determined distances for 940 molecular clouds by matching them with the clump and cloud catalogs of \cite{Duarte2021-500, Urquhart2022-510}. For the 364 clouds that did not have matches in the catalogs of \cite{Duarte2021-500, Urquhart2022-510}, in this work, we further cross-matched them with the cloud catalog of \cite{Miville2017-834} to assign distances. For the remaining subset of clouds that still lacked matches, we estimated their kinematic distances following the method of \cite{Miville2017-834}, using the Galactic rotation curve of \cite{Brand1993-275}. 

As an extension toward smaller spatial scales, we further decompose molecular clouds in the Milky Way into smaller substructures using the dendrogram algorithm, based on the same dataset presented in \cite{bridge}. Here, we require that each substructure has a minimum area larger than 10 beam sizes, since structures that are too small may invalidate the methods used in equations \ref{21um} and \ref{mass} to estimate the mass and the SFR. 
Additionally, we adopt the mean intensity of each molecular cloud as the threshold, which means that we only extract the regions within each cloud with intensities above the mean.
As an extension toward larger spatial scales, we examine nearby galaxies observed with single-dish telescopes. We select several galaxies from the HERACLES (Heterodyne Receiver Array CO Line Extragalactic Survey) survey (\cite{Leroy2009-137}, with a resolution of 13$''$) that are already included in Table.S\ref{near}—namely NGC 0628, NGC 4321, NGC 4254, NGC 3351 and NGC 3627—and identify structures within these galaxies.
Finally, we obtained 6835 molecular cloud substructures and 415 kpc-scale molecular gas structures.

As shown in Fig.S\ref{s-multi}, molecular clouds in the Milky Way and their internal substructures, as well as gas structures identified on different spatial scales in the same galaxies using ALMA and single-dish observations, all follow a similar $M_{\rm H_{2}}$--SFR relation. This demonstrates that the observed $M_{\rm H_{2}}$--SFR relation is robust against different choices of spatial scales and structure identification methods.
Within our clump-based framework, this scale invariance can be naturally understood as a consequence of larger-scale structures representing ensembles of star-forming clumps. Their differences reflect variations in the number of embedded clumps, causing them to move along the $M_{\rm H_{2}}$--SFR relation rather than systematically deviate from it. 
Therefore, the observed consistency across scales provides an important constraint for models of star formation and can be interpreted within a framework in which cloud-scale star formation emerges from the collective activity of clump populations.

\subsection{Molecular gas in the entire galaxy}

The lowest-resolution or largest-scale structure is obtained by treating the entire galaxy as a single entity. We compile tables of galaxy physical parameters from the literature, with a particular focus on catalogs containing dwarf galaxies or low-metallicity systems. The samples in these tables are drawn from various nearby galaxy surveys \cite{Wyder2009-696,Schruba2012-143,Aniano2020-889,Hunt2019-621,Leroy2021-257,Stark2013-769,Kewley2005-117,Hunt2015-583,Hunt2017-606,Cormier2014-564,Grossi2016-590,Schruba2019-883,delosReyes2019-872}. The final sample consists of 311 unique galaxies. These galaxies span a wide range of types and cover a broad parameter space, as shown in Fig.S\ref{s-para}.
In these samples, the CO-to-H$_{2}$ conversion factor used to derive molecular gas masses has been corrected for metallicity, although the adopted methods are not uniform across the literature. In addition, the methods used to estimate the SFR are also not uniform. These measurement inconsistencies likely contribute significantly to the scatter of the data points. Nevertheless, a clear $M_{\rm H_{2}}$--SFR relation can still be observed, which is broadly consistent with that seen at smaller spatial scales, as shown in Fig.S\ref{s-all}. 
Furthermore, we divide the sample into two subsamples based on solar metallicity (Fig.S\ref{s-para}) and find that the $M_{\rm H_{2}}$–SFR relations of high- and low-metallicity galaxies are comparable.

\subsection{Molecular clouds in the outer Milky Way}\label{outer}

About 95\% of molecular clouds in the Milky Way listed in Table.S\ref{near} have Galactocentric distances smaller than 10 kpc.
To further increase the fraction of outer-Galaxy molecular clouds, we identify 1344 molecular clouds from the $^{12}$CO (1-0) data of the MWISP (Milky Way Imaging Scroll Painting, $10^\circ \le l \le 230^\circ$ and $-5.25^\circ \le b \le 5.25^\circ$) survey \cite{Yang2026-282} using the same method as in \cite{bridge}, and determine their distances using the methods described above (Sec.\ref{sub-pc}). 
Among them, 47\% of the molecular clouds have Galactocentric distances larger than 10 kpc. 
For molecular clouds in this sample not covered by \textit{Spitzer} observations, we estimate their SFRs using mid-infrared images (21.3 $\mu$m) from Midcourse Space Experiment (MSX) \cite{Price2001-121}. As shown in Fig.S\ref{s-outer}(a), compared to molecular clouds in the inner Milky Way, those located in the outer Milky Way exhibit systematically higher SFRs. Given that the outer Milky Way also has significantly lower metallicity \cite{Bolatto2013-51,Lian2023-7}, molecular clouds in this region are analogous to those in low-metallicity dwarf galaxies (Fig.S\ref{s-outer}(b)), where the scatter of the $M_{\rm H_{2}}$--SFR relation is large. This strengthens our argument that molecular clouds in the Milky Way can serve as useful templates for studying molecular clouds in nearby galaxies. However, the systematically higher SFRs inferred for the outer-Galaxy clouds may also be affected by the sensitivity and noise properties of the MSX data, and whether this trend is physical requires further investigation. Moreover, 
due to the dependence of the CO-to-H$_{2}$ conversion factor on metallicity, the masses of these outer-Galaxy molecular clouds cannot be well estimated. Therefore, we do not discuss this sample further; see Sec.\ref{h-sfr} for more details.

\section{Molecular clouds with higher SFRs}\label{h-sfr}

As shown in Fig.S\ref{s-all}, the $M_{\rm H_{2}}$--SFR relation is broadly comparable for dwarf and disk galaxies, both for sub-galactic structures and for entire galaxies. However, a subset of molecular clouds or molecular cloud complexes in dwarf galaxies appear to exhibit higher SFRs.
Several possible explanations may account for this:

(1) The criteria $r_{\rm m} >1$  and $r_{\rm p} >1$ introduced in Sec.\ref{calculation} may not fully exclude highly evolved molecular clouds. The structures identified as molecular clouds may in fact be remnants after stellar feedback, leading to a significant underestimation of their original gas masses;

(2) The metallicity of individual molecular clouds in each galaxy is not known. In order to enable a direct comparison with disk galaxies, we do not apply metallicity-dependent corrections to the CO-to-H$_{2}$ conversion factor for the spatially resolved structures within dwarf galaxies. This is also due to the substantial uncertainties associated with such corrections. 
In particular, the conversion factor does not solely depend on the metallicity \cite{Hunter2024-62}. For example, NGC 6822 has a significantly lower metallicity than the LMC, yet appears to require a smaller correction to the conversion factor \cite{Schruba2017-835,Hunter2024-62}. 
In addition, the metallicities of M33 and LMC are comparable and significantly lower than that of the Milky Way, but the conversion factor of M33 is approximately at or below the Galactic value \cite{Keilmann2024-692}.
In the literature, the CO-to-H$_{2}$ for molecular clouds in the LMC is reported to be $\sim$2–10 times higher than the Galactic value \cite{Fukui2008-178,Roman-Duval2014-797}, and for 30 Doradus it can reach $\sim$10–50 times higher \cite{Fukui2008-178,Chevance2020-494}, depending on the estimation methods. Therefore, applying a single, uniform metallicity-dependent correction factor to all molecular clouds within a galaxy may not be appropriate.
Nevertheless, it is well established that the conversion factor should be significantly larger in low-metallicity dwarf galaxies than in typical disk galaxies. As a result, we are underestimating the gas masses of some molecular clouds in dwarf galaxies, which can artificially enhance their apparent SFRs by shifting them horizontally to the left on the $M_{\rm H_{2}}$--SFR plane.

(3) The above effects can be interpreted as systematic underestimation of molecular gas masses, which in turn produces apparent deviations from the $M_{\rm H_{2}}$--SFR relation.
Alternatively, the elevated SFRs may be intrinsic in some cases, for example in certain molecular clouds in the LMC that are undergoing a starburst phase driven by the interaction between the LMC and the Milky Way. 
In such cases, a Gaussian star formation history (SFH) should be more appropriate than a flat SFH. 
In \cite{model}, we defined two types of SFH: flat and Gaussian. The Gaussian SFH is specifically used to characterize molecular clouds that undergo a starburst phase, or, more generally, molecular clouds whose star formation activity varies with time.
As predicted by the model, using a Gaussian SFH naturally yields higher instantaneous SFRs than a flat SFH, as it implies a higher number of star-forming clumps.

A possible physical explanation for the enhanced SFRs observed in low-metallicity environments is provided by feedback-regulated star formation models. In particular, studies by \cite{Dib2011-415,Dib2011-737} suggest that stellar feedback from massive OB stars is strongly metallicity-dependent. At lower metallicities, stellar winds are weaker, which reduces their ability to expel gas from protocluster clumps. As a result, star formation can proceed over longer timescales, leading to higher final star formation efficiencies (SFEs). For molecular clumps with otherwise identical properties (e.g., mass and density profile), this implies that lower-metallicity environments naturally sustain more prolonged and efficient star formation episodes. This mechanism provides a plausible physical origin for the elevated SFRs, complementing phenomenological descriptions based on bursty SFHs.

Additionally, these molecular clouds may also have a significantly higher fraction of dense gas, which would increase the SFRs, as discussed in \cite{Lada2012-745}. In our model, more dense gas implies more star-forming clumps, so the overall SFR will be higher.
As shown in Fig.1 of \cite{model}, the dense gas fraction varies significantly among different molecular clouds, while in our model fitting we adopt only an average dense gas fraction. 
Beyond the dense gas fraction, variations in the SFR also depend on the evolutionary stage of molecular clouds, which is reflected in differences in their SFHs. In our fitting, we use a flat SFH to approximate the overall distribution. Actually, individual molecular clouds have different ages and are at different evolutionary stages, and therefore possess their own SFHs. For a large population of clouds spanning a range of ages, the mixture and averaging of their diverse SFHs can be reasonably approximated by a flat SFH. 
This way, we are only fitting the overall distribution of all molecular clouds, without accounting for some extreme cases.

In summary, regardless of the underlying cause, these high-SFR molecular clouds remain consistent with the theoretical framework adopted in this work and do not fall outside its scope. 
Moreover, as shown in Fig.\ref{s-outer}(b),
molecular clouds in dwarf galaxies are broadly similar to those in the outer Galaxy; therefore, detailed investigations of the latter can shed light on the former and help discriminate among the scenarios outlined above.

To draw a conclusion,
we use the 99\% prediction interval to characterize the scatter in the $M_{\rm H_{2}}$--SFR relation for Milky Way clouds in Table.S\ref{near}. We then identify molecular clouds in three dwarf galaxies (LMC, SMC and NGC 6822) with significantly elevated SFRs as those lying above the upper boundary of this prediction interval. These clouds account for $\sim$15\% of the clouds in these three dwarf galaxies.
Based on the analysis above, the molecular cloud masses in these low-metallicity dwarf galaxies are likely underestimated. To shift these clouds down to the upper boundary of the prediction interval, their masses would need to increase by a median factor of $\sim$2.1. Bringing them fully within the prediction interval would require larger correction factors. These values are comparable to correction factors inferred from independent observational methods. Therefore, this result supports the interpretation that the apparent offset from the $M_{\rm H_{2}}$--SFR relation of disk galaxies is caused by underestimated molecular cloud masses in low-metallicity dwarf galaxies.

As shown in Fig.S\ref{s-multi}(a), some substructures within Milky Way molecular clouds also exhibit elevated SFRs similar to those observed in LMC clouds. Compared to entire molecular clouds, these substructures contain less diffuse gas because of the threshold applied in the analysis (Sec.\ref{sub-pc}), which effectively reduces the inferred total gas mass. This effect is, to some extent, analogous to the situation in low-metallicity environments, where CO is preferentially destroyed in low-density regions. As a result, only the denser gas components remain effectively traced by CO, leading to apparently elevated SFRs at a given CO-inferred molecular gas mass, as seen for both the Milky Way substructures and the LMC clouds in Fig.S\ref{s-multi}(a).

\section{Scatter and galactic environment}\label{environment}

\begin{figure*}
\centering
\includegraphics[page=1,width=0.9\textwidth]{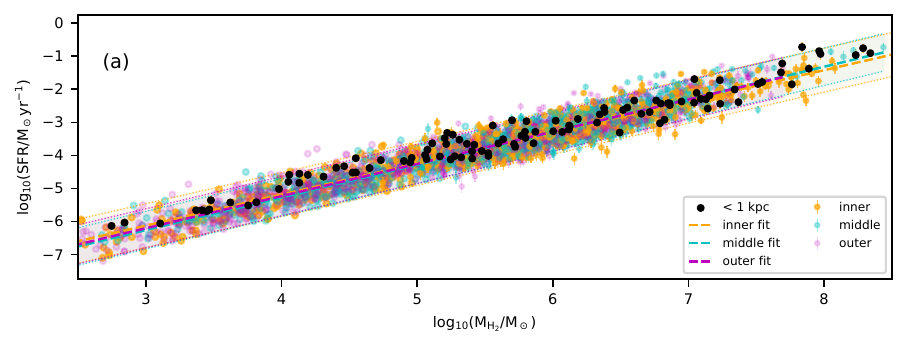}

\vspace{-25pt}
\includegraphics[page=2,width=0.9\textwidth]{fig/env.pdf}

\vspace{-15pt}
\caption{
The $M_{\rm H_{2}}$--SFR relation in different galactic environments. 
(a) Molecular clouds in each galaxy are divided into four groups based on their galactocentric distance (inner, middle, outer, within 1 kpc); 
(b) Molecular clouds in each galaxy are divided into three groups based on their molecular gas surface density (high, middle, low).
For each group, the fit was performed using unweighted ordinary least squares. The 95\% prediction bands are shown as shaded regions with dashed boundary lines.
}
\label{env}
\end{figure*}

The consistency emphasized in this work is based on the result that molecular gas structures on different scales follow a similar $M_{\rm H_{2}}$--SFR relation. However, this relation exhibits significant scatter, which likely arises from multiple factors, such as those discussed above (Sec.\ref{h-sfr}).
The physical properties of molecular clouds vary across different regions within the same galaxy and also differ from one galaxy to another \cite{Querejeta2021-656,Kim2022-516,Sun2022-164}, therefore, the galactic environment may also be a contributing factor to the scatter.
In such works, the molecular gas surface density is often used as a proxy for the local galactic environment. Following this approach, we divide the molecular clouds in each galaxy into three groups based on their molecular gas surface density.
In Table.S\ref{near}, only the Milky Way and 16 nearby galaxies have well-defined galactic centers. 
Since the environments in galactic centers and outskirts are generally different, we also divide the molecular clouds in each galaxy into three equal-sized groups according to their galactocentric distances. In addition, we define a separate group consisting only of molecular clouds located within the central 1 kpc of the galaxies.
For all these cases, as shown in Fig.S\ref{env}, we do not find significant differences in the $M_{\rm H_{2}}$--SFR relation.
Nevertheless, the galactic environment may still contribute to the scatter in the $M_{\rm H_{2}}$--SFR relation, for example by altering the dense gas fraction. In the studies mentioned above, the HCN-to-CO ratio is used as a tracer of the dense gas fraction, and this ratio is found to vary with galactic environment. However, whether HCN reliably traces dense gas is currently under debate \cite{Pety2017-599,Priestley2026-547}.
In addition, the dense gas fraction may also vary with the evolutionary stage of molecular clouds.

\section{Model parameters}\label{procedure}

\begin{figure*}
\centering
\includegraphics[page=1,width=0.9\textwidth]{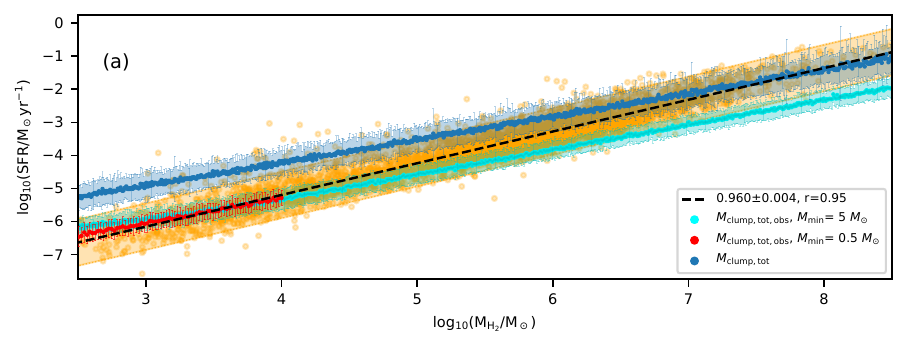}

\vspace{-25pt}
\includegraphics[page=2,width=0.9\textwidth]{fig/test.pdf}

\vspace{-15pt}
\caption{
Background orange points represent molecular clouds in 30 galaxies listed in Table.S\ref{near}. 
The black dashed line shows the linear fit (unweighted ordinary least squares) to the orange points.
$r$ is the correlation coefficient. 
The 95\% prediction bands are shown as shaded regions with dashed boundary lines.
(a) The blue and cyan lines show the predictions based on equation.\ref{Mcti} and equation.\ref{Mcto}, respectively, with the shaded regions indicating the corresponding uncertainty ranges. Both the red and cyan lines are based on equation.\ref{Mcto}, but the minimum clump mass adopted in the model is 0.5 $M_{\odot}$ and 5 $M_{\odot}$, respectively.
(b) Predictions based on equation.\ref{Mctio} for different values of $a$ and $b$. The vertical bars indicate the uncertainty ranges. Due to the similarity of the fitted lines, overlap is inevitable.}
\label{test}
\end{figure*}

In Fig.\ref{msfr}, we fit the observations using a theoretical model, considering only the sample listed in Table.S\ref{near}. 
The construction and implementation of the model are described in detail in \cite{model}.

Clumps are identified as the primary sites of star formation within molecular clouds. The SFR on molecular cloud scales, as a large-scale phenomenon, is in fact the integrated manifestation of star formation processes occurring at the clump scale within molecular clouds.
For molecular clouds in the Milky Way, their internal clumps can be resolved, allowing us to quantify the relationship between molecular clouds and their embedded clumps; this forms the foundation of our model. The model primarily provides a theoretical framework that integrates a wide range of observational results, thereby clarifying priorities for future observations.
Unlike previous theoretical models \cite{Krumholz2005-630,Padoan2011-730,Hennebelle2011-743,Federrath2012-761,Zamora2014-793,Vazquez2019-490,Ostriker2011-731,Semenov2017-845,Kruijssen2019-569}, this work adopts a predominantly observation-driven approach. Rather than relying on specific theoretical assumptions, we attempt to synthesize existing observational results and treat the molecular cloud-scale SFR as a natural consequence of a set of empirical relations.
Turbulence, gravity, feedback, magnetic fields, evolutionary stage, and other processes all contribute to the observed results; however, we do not attempt to disentangle their individual roles, and instead focus solely on the observational outcomes.
The reliability of the model is therefore grounded in the robustness of the observations. As such, it can be readily extended and updated as new observations become available, allowing the model to improve alongside advances in observational data.

Our model is observationally driven rather than being based on an ab initio theory of star formation.
We view this as a complementary approach to ab initio theories rather than an alternative to them. We seek to integrate a broad range of empirically established relations into a unified framework, while recognizing that the present model does not explain their fundamental physical origin.
Nevertheless, these empirical relations are robust observational results that any successful theory of star formation must ultimately account for. The physical mechanisms responsible for them—including the relative roles of gravity, turbulence, magnetic fields, stellar feedback, and environmental effects—remain the subject of active investigation, and different theoretical models offer different interpretations. By remaining independent of any specific theoretical scenario, our framework provides an observationally constrained description of the cloud-scale star formation relation. Establishing a direct connection between this empirical framework and first-principles (ab initio) models is an important direction for future work.

In our previous work, we applied the model to interpret the observed $M_{\rm H_{2}}$--SFR relation in Milky Way molecular clouds.
In this work, we extend this approach to explore the consistency of star formation across the Milky Way and nearby galaxies. By comparing Milky Way molecular clouds with those in a large sample of nearby galaxies of different types and environments, we find that they follow a similar $M_{\rm H_{2}}$--SFR relation. This suggests that the model can also be applied to molecular clouds in nearby galaxies.

The main source of uncertainty in the model arises from the estimate of the total initial clump mass in molecular clouds. 
For observations in the Milky Way, the total mass of clumps currently present in a molecular cloud ($M_{\mathrm{clump,tot,obs}}$) is related to the cloud mass as follows
\begin{equation}
\begin{split}
\mathrm{log}_{\rm 10} (M_{\mathrm{clump,tot,obs}}/M_{\odot}) =  
(0.75 \pm 0.02) \times \mathrm{log}_{\rm 10} (M_{\mathrm{cloud}}/M_{\odot}) - (0.56 \pm 0.13).
\end{split}
\label{Mcto}
\end{equation}

The molecular clouds we observe could be composites, containing both
exposed and embedded stellar populations, as well as clumps. Nearly all of those stellar populations—whether visible or still embedded—originate from clumps. These clumps may either be currently existing or have existed in the past. Past clumps are now observed as exposed and embedded stellar populations. 
Together, existing and past clumps constitute the initial clump population of the cloud, described by the initial clump mass function defined in \cite{model}. 
As an estimate, the total initial clump mass is
\begin{equation}
M_{\mathrm{clump,tot}} 
= M_{\mathrm{clump,tot,obs}} +
   M_{\mathrm{*,tot,obs}}/\mathrm{SFE_{\mathrm{clump}}},
\label{Mcti}
\end{equation}
where $M_{\mathrm{*,tot,obs}}$ is the total mass of the stellar population contained within the molecular cloud. 
The clump star formation efficiency ($\mathrm{SFE}_{\mathrm{clump}}$), with a median value of $\approx 0.3$, decreases with increasing clump mass \cite{Zhou2024PASP-2}.
Adopting a median value of 0.3 would therefore lead to an overestimate of the total clump mass.
For clouds in nearby galaxies with masses larger than 
$10^{5.5}$ $M_{\odot}$,
in \cite{Zhou2025-541}, we derived a relation 
\begin{equation}
\begin{split}
\mathrm{log}_{\rm 10} (M_{\mathrm{*,tot,obs}}/M_{\odot}) = 
(0.68 \pm 0.04) \times \mathrm{log}_{\rm 10} (M_{\mathrm{cloud}}/M_{\odot}) + (0.19 \pm 0.27).
\end{split}
\label{Mst}
\end{equation}
Same as the findings of \cite{model}, in Fig.S\ref{test}, equation.\ref{Mcti} mainly fits the high-mass end of molecular clouds, while equation.\ref{Mcto} provides a good fit for the low-mass end. 
For molecular clouds with masses less than 10$^{3}$ $M_{\odot}$, the deviation of the model is mainly due to the minimum clump mass ($M_{\mathrm{min}}$) being restricted to 5 $M_{\odot}$ \cite{model}. If the minimum clump mass for sampling is reduced to 0.5 $M_{\odot}$, the model performs better. The mass threshold can be further lowered to 0.1 $M_{\odot}$ or even smaller. In this way, dense cores can be naturally included in the model.

Physically, molecular clouds at the high-mass end can survive for a long time despite hosting stellar populations. In contrast, in low-mass molecular clouds, feedback from the stellar populations can efficiently disrupt the cloud. Hence, observed low-mass molecular clouds are not expected to contain a significant fraction of stellar components. Naturally, as the molecular cloud mass increases, its ability to form and retain stellar populations also becomes stronger. 
In addition, equation.\ref{Mcti} also overestimates the total mass of clumps in practice. 
Therefore, we need a new relation: it should follow equation.\ref{Mcti} for high-mass molecular clouds, equation.\ref{Mcto} for low-mass molecular clouds, and lie between the predictions of the two equations for intermediate-mass molecular clouds.
As done in \cite{model}, to ensure a smooth transition,
we combine equation.\ref{Mcto} and equation.\ref{Mcti} using a linear weighting function defined as
\begin{equation}
t = \frac{\mathrm{log}_{\rm 10} (M_{\mathrm{cloud}}/M_{\odot}) - a}{b - a}, 
\end{equation}
where $a$ and $b$ specify the transition mass range, and 
$t$ is limited to the range [0,1].
The final relation is then given by
\begin{equation}
M_{\mathrm{clump,tot,m}} = (1 - t)\, M_{\mathrm{clump,tot,obs}} + t\, M_{\mathrm{clump,tot}},
\label{Mctio}
\end{equation}
where $M_{\mathrm{clump,tot,obs}}$ and $M_{\mathrm{clump,tot}}$ correspond to the relations of 
equation.\ref{Mcto} and equation.\ref{Mcti}, respectively. 
In \cite{model}, using observations in the Milky Way as a reference, we estimate the optimal values of $a$ and $b$ to be 4 and 6.5, respectively. 
The combination of equation.\ref{Mcto} and equation.\ref{Mcti} partially compensates for the overestimation of the total clump mass caused by adopting a value of 0.3 for $\mathrm{SFE}_{\mathrm{clump}}$, thereby also contributing to a better fit to the observational data. However, the model predictions are still slightly overestimated.
As a comparison, we also explore the cases $a$=4.5, $b$=7 and $a$=3.5, $b$=6. As shown in Fig.S\ref{test}, all yield very similar results in terms of slope and intercept.
Nevertheless, the values of $a$ and $b$ should be further constrained observationally. 
A more advanced step in improving the model is to empirically derive the quantitative relationship between molecular cloud mass and the mass of their internal stellar populations across the full range of cloud masses.

In Fig.S\ref{test}, we estimated the best-fitting linear relation using ordinary least-squares (OLS) regression. The shaded region represents the 95\% prediction interval, which was calculated from the residual variance of the fit and therefore reflects the expected dispersion of individual data points around the best-fitting relation.
The prediction interval reflects the scatter of the data about the fitted relation but does not explicitly include systematic uncertainties associated with the CO-to-H$_{2}$ conversion factor, SFR calibrations, distance estimates, or correlated uncertainties.
No measurement uncertainties were used as weights in the fitting procedure. Consequently, the quoted uncertainties describe the statistical dispersion of the observed sample rather than a propagation of individual measurement errors.
Many systematic uncertainties are difficult to quantify robustly and may themselves change as observational techniques improve. Rather than attempting to account for every uncertain systematic effect, we focus on whether the model reproduces the overall observed relation within the statistical dispersion of the data. The good agreement in both slope and intercept between the model and the observations, despite these unavoidable uncertainties, demonstrates the effectiveness of the proposed framework. As we noted, the model should evolve together with improvements in observational constraints as increasingly accurate measurements become available.

\end{document}